\documentclass[11pt,a4paper]{article}

\usepackage{amsmath}
\usepackage{amssymb}
\usepackage{graphicx}
\usepackage{cite}
\usepackage{hyperref}
\usepackage{enumerate}
\usepackage{longtable}
\usepackage{titlesec}

\titleformat{\chapter}[display]
{\normalfont\bfseries}{}{0pt}{\Huge}

\usepackage{xcolor}
\definecolor{bl}{rgb}{0.0,0.2,0.6}

\hypersetup{
    unicode=true,           %
    pdftoolbar=false,       %
    pdfmenubar=false,       %
    pdffitwindow=false,     %
    pdfstartview={FitH},    
    pdftitle={},            
    pdfauthor={},           
    pdfsubject={},          
    pdfcreator={},          
    pdfproducer={},         
    pdfkeywords={},         
    pdfnewwindow=true,      
    colorlinks=true,        
    linkcolor=bl,           
    citecolor=red,          
    filecolor=green,        
    urlcolor=red            
}

.tfm scaled 1000 
.tfm scaled 1000 
.tfm scaled 1000 

\begin{document}
\title{\sf Single [2]catenanes in solution forming ${\sf 4}$-plats: a combined field theoretical and numerical approach}

\author{
Franco Ferrari\footnote{e-mail: franco.ferrari@usz.edu.pl}
$\quad\quad$
Marcin R. Pi\c{a}tek\footnote{e-mail: marcin.piatek@usz.edu.pl}
\\[8pt]
CASA* and Institute of Physics, University of Szczecin\\ 
Wielkopolska 15, 70--451 Szczecin, Poland}
\date{}


\maketitle

\begin{abstract}\noindent
The statistical mechanics of [2]catenanes in a solution with
constrained number of  maxima and minima along a special direction (the height) is discussed. The interest in this system comes 
from the fact that, in the homopolymer case, it has analogies with self-dual anyon field theory models and has conformations
that minimize the static energy and bear particular symmetries and properties.
In the first part of this work we provide a procedure for deriving the equations of motion
in the case of replica field theories in the  limit of zero replicas. We compute also the partition function of the [2]catenane 
at the lowest order in the frame of the so-called background field method.
In the second part the statistical mechanics of the system is investigated using numerical simulations based on the Wang-Landau 
Monte Carlo method.
At equilibrium, independently of the temperature, it turns out that the conformations of the system are elongated in the height 
directions. The two rings composing the [2]catenanes have approximately the same heights and are aligned.
The thermodynamic properties of the system are discussed and the results coming from the field theoretical approach are compared 
with those of the numerical simulations.  
\end{abstract}

\newpage
\tableofcontents 

\section{Introduction}\label{introd}
$2s-$plats, with $s=1,2,\ldots$, are structures formed by interlocked polymer rings. These structures are  characterized by a fixed number $s$ of maxima and a fixed number $s$ of minima in one special direction, which we chose here to be the height $t$ of the system. The use of the variable $t$ instead of the variable $z$ is due to the analogy between the statistical mechanics of $2s-$plats with a system of $2s$ anyons established in
\cite{FF2004,FFJPMPYZ2019}, where there are two spatial dimensions and the third dimension is time.
The constraint of fixed number of maxima and minima  implies that
the polymer lines going from any point of  minimum to the nearest point of maximum are monotonic curves. More rigorously, in mathematics $2s-$plats are defined as closures of a number of $2s$ braids.
The particular case $s=2$ of $4-$plats investigated in this paper is common in DNA, see e.g.~\cite{ernst1,ernst2,kaufmann1,sumners1,Vazquez,Vazquez2}.

The present work focuses on $4-$plats formed by [2]catenanes, i.e., two rings linked together.  Systems of this type have been thoroughly investigated in the past, for instance in \cite{fujita,wassermann1,wassermann2,Alex1,Alex2,Mesfin,Eutopiareview,Rensburg}.
The statistical properties of $4-$plats may be described
using a model of replica complex scalar fields \cite{FF2004} that has been later generalized  in \cite{FFJPMPYZ2019} by adding interactions and allowing for general values of $s$. Roughly speaking, the scalar fields are related to the monomer densities of the polymer lines.
In the field theoretical approach considered here, the interactions are switched off, but there are still the interactions of entropic origin due to the topological constraints.  The latter are imposed using the Gauss linking number, one of the simplest topological invariants that can be applied to distinguish the topological state of  links formed by two rings. The Gauss linking number can be written as the flux of a magnetic field across the surface spanned by one of the rings.
The magnetic field is generated by a fictitious current flowing along the other ring.
In summary, in the field theoretical model the interactions of topological origin between the monomers, or, better, the complex scalar fields, are mediated by magnetic fields. Of course, magnetic fields are connected with long-range forces.
On the other side, the topological constraints forbid that the polymer lines cross themselves. This requires also short-term interactions. Accordingly, after the passage from polymer paths to fields is completed, in the action of the field theoretical model appear also terms that take into account short-range Coulomb-like interactions.
 The part of the action responsible for the long-range interactions has
 a property that is called self-duality \cite{Dunne}. The terms responsible for the short-range interactions, which are not self-dual, are shown to vanish when the rings are homopolymers \cite{FF2004}.
 
The main motivation of this study comes from Ref. \cite{NAFFMP2024}.
In particular, in \cite{NAFFMP2024} two classes of self-dual solutions of the field equations of motion have been derived. These solutions minimize
the static energy of the $4-$plat and are invariant under  symmetry  transformations: For instance, a large classe of solutions with conformal symmetry can be found assuming that the height $T$ of the [2]catenane is large and that, at each given height, there are no differences in the monomer densities of the polymer lines \footnote{In the average, in a scale of time that is large with respect to the time scale of thermal fluctuations, this assumption is true.}.
Other solutions with translational symmetry in one of the directions
on the plane that is perpendicular to the height, have been obtained in the case in which the dimension of the system in the other direction is small.

Of course, self-dual solutions are usually found in spin systems. Polymers are instead related to the universality class of systems with spin zero.
To solve this apparent contradiction, we derive here the self-dual equations of motion in a way that shows explicitly that the first replica fields, corresponding to a spin 1 field theory, have non-vanishing degrees of freedom. In doing that, we use the fact that the probability distribution measuring the probability that the two rings wind up one around the other in such a way that the required topological constraints are fulfilled,  is formally written as a field theory propagator of the first replica fields.
In the following, the probability distribution will be referred as partition function.

Further, the possibility of using the background field method~\cite{abbott} is considered. Within this approach, the partition function is computed starting as a zeroth approximation from a classical background field configuration.  The corrections due to the thermal fluctuations are then evaluated by means of perturbation theory. In this work we
provide the explicit form of the partition function at the zeroth order using as a background a generic solution of the self-dual equations of motion.
Of course, this strategy works at its best if the solutions of the field equations
correspond to a deep point of minimum in the energy landscape.
However, the analytical calculations of \cite{NAFFMP2024} point out that the
energy landscape of the $4-$plat  is rather complex being characterized by a plethora
of points of minimum. The depth of the minima cannot be estimated by analytical methods, so we have switched to numerical simulations.
Such simulations allow to evaluate the free energy of the system.
Moreover, we have searched for possible symmetries emerging in the sampled conformations when the system is at equilibrium. The goal is to check if some of the symmetries predicted by the analytical model are still present when the thermal fluctuations are acting.
The sampling of the statistically relevant conformations of the $4-$plat on a simple cubic lattice is performed by means of the Wang-Landay Monte Carlo algorithm \cite{wl}.
With respect to the theoretical model, the numerical approach presents two differences.
In particular, the system is subjected to short-term attractive interactions. Despite that, the comparison with the field theory is still possible in the limit of high temperatures, in which the interactions become irrelevant due to the strong thermal fluctuations.
Secondly, the heights of the two rings are not constrained to be equal to some large value $T$ as in the field theoretical approach, but are able to fluctuate. Remarkably, in all inspected conformations the final heights of the two rings are almost the same and are large with respect to the other dimensions of the [2]catenane as assumed in the field theoretical model.
The other results obtained in this paper will be discussed in the Conclusions.

The material presented in this work is divided as follows. In Section \ref{ftapproach} the field theoretical approach to the statistical mechanics of $4-$plats is discussed. The self-dual equations of motion of the system are derived together with the expression of the partition function of the system computed in the background of a self-dual solution.
The results of the numerical simulations are the subject of Section~\ref{numsim}.
Finally, the conclusions are drawn in Section~\ref{conclusions}.
\section{Field theoretical approach}\label{ftapproach}
The starting point is the following partition function:
\begin{equation}
  Z(\lambda,\alpha)=\int{\cal D}(fields){\rm e}^{-S_{matter}}{\rm e}^{\alpha\prod_{a=1}^2
    \psi_a^{u(1)*}(\boldsymbol r_{a,2},T)    \psi_a^{u(1)}(\boldsymbol r_{a,1},0)
    \psi_a^{d(1)*}(\boldsymbol r_{a,1},0)    \psi_a^{d(1)}(\boldsymbol r_{a,2},T)},\label{pint}
\end{equation}
where $\alpha$ is a small parameter and
\begin{eqnarray}
  {\cal D} (fields)&=&\prod_{a=1}^2\left[
    {\cal D}\psi_a^{u(1)*}(\boldsymbol x,t) {\cal D}\psi_a^{u(1)}(\boldsymbol x,t)
    \cdots
    {\cal D}\psi_a^{u(n)*}(\boldsymbol x,t)    {\cal D}\psi_a^{u(n)}(\boldsymbol x,t)
    \right]\nonumber\\
  &&\prod_{a=1}^2
    \left[{\cal D}\psi_a^{d(1)*}(\boldsymbol x,t) {\cal D}\psi_a^{d(1)}(\boldsymbol x,t)
    \cdots
    {\cal D}\psi_a^{d(n)*}(\boldsymbol x,t)    {\cal D}\psi_a^{d(n)}(\boldsymbol x,t)
    \right].
\end{eqnarray}
Moreover, ${  \vec \Psi}^{u*},{  \vec \Psi}^{u},{  \vec \Psi}^{d*}, {  \vec \Psi}^{d}$
are replica complex scalar fields:
\begin{equation}
  {  \vec \Psi}^{u,d*}(\boldsymbol x,t)=
  \left(\psi^{u,d(1)*}(\boldsymbol x,t),\ldots,\psi^{u,d(n)*}(\boldsymbol x,t)
  \right),
\end{equation}
\begin{equation}
    {  \vec \Psi}^{u,d}(\boldsymbol x,t)=
  \left(\psi^{u,d(1)}(\boldsymbol x,t),\ldots,\psi^{u,d(n)}(\boldsymbol x,t)
\right).
\end{equation}
The superscript $u,d$ means $u$ albo $d$ fields. The symbol $\left|{\vec \Psi}^{u,d}\right|^2$ denotes the scalar product over 
replica indices: $\left|{\vec \Psi}^{u,d}\right|^2=\sum_{\rho=1}^n\psi^{u,d(\rho)*}\psi^{u,d(\rho)}$.
Here, $(\boldsymbol x,t)$, where $\boldsymbol x=(x^1,x^2)$,
is a point in a slice of a three-dimensional space in which
the height $t$ varies within the interval $[0,T]$.
Middle latin letters $i,j,\ldots=1,2$ will be used for the 2d spatial indices. 
The partial derivatives $\frac{\partial}{\partial x^i}$ and $\frac{\partial}{\partial t}$
will be denoted $\partial_i$ and $\partial_0$ respectively.

The matter action $S_{matter}$ is splitted into three parts:
\begin{equation}
S_{matter}=I_{sd}+I_C+I_T.
\end{equation}
$I_{sd}$ contains the self dual contribution:
\begin{equation}
  I_{sd}=\sum_{a=1}^2\int_0^Tdt\int d^2x\left[
\frac 1{4g_{a,u}}|(D_{a,1}^u+iD_{a,2}^u){\vec\Psi}_a^u|^2
+\frac 1{4g_{a,d}}|(D_{a,1}^d+iD_{a,2}^d){\vec\Psi}_a^d|^2
    \right].\label{Isd}
\end{equation}
$I_C$ takes into account short-range Coulomb-like interaction:
\begin{eqnarray}
  I_C&=&\frac\lambda{8\pi}\int_0^Tdt\int d^2x\left[
      \left(-\frac 1{g_{1,u}}|{\vec \Psi^u_1}|^2+\frac 1{g_{1,d}}|{\vec \Psi^d_1}|^2
      \right)
      \left(-|{\vec \Psi^u_2}|^2+|{\vec \Psi^d_2}|^2
      \right)
      \right.
      \nonumber\\
      &+&
      \left.
      \left(\frac 1{g_{2,u}}|{\vec \Psi^u_2}|^2-\frac 1{g_{2,d}}|{\vec \Psi^d_2}|^2
      \right)
      \left(-|{\vec \Psi^u_1}|^2+|{\vec \Psi^d_1}|^2
      \right)
      \right].\label{Ic}
\end{eqnarray}
Finally, the terms with the derivatives $\partial_0$ with respect to the height $t$ are confined in $I_T$:
\begin{equation}
  I_T=\sum_{a=1}^2\int_0^Tdt\int d^2x\left[
\vec\Psi_a^{u*}\partial_0\vec\Psi_a^u+\vec\Psi_a^{d*}\partial_0\vec\Psi_a^d
    \right].\label{IT}
\end{equation}
In Eq.~(\ref{Isd}) the symbols $D_{a,i}^{u,d}$ denote the covariant derivatives:
\begin{equation}
  D_{a,j}^u=\partial_j-i\sum_{b=1}^2c_{ab}B_{b,j},\qquad\qquad
  D_{a,j}^d=\partial_j+i\sum_{b=1}^2c_{ab}B_{b,j},
\end{equation}
where the magnetic fields $\boldsymbol B_b(\boldsymbol x,t)$ are the solutions of the equations:
\begin{eqnarray}
  \frac 1{4\pi}\epsilon^{ij}\partial_iB_{1,j}&=&c_{12}(-|\vec \Psi_1^u|^2 +|\vec \Psi_1^d|^2 ),
  \\
  \frac 1{4\pi}\epsilon^{ij}\partial_iB_{2,j}&=&c_{21}(-|\vec \Psi_2^u|^2 +|\vec \Psi_2^d|^2 ),
\end{eqnarray}
$c_{ab}$ is a $2\times 2$ matrix of coupling constants:
\begin{equation}
  c_{ab}=
    \begin{bmatrix}
      0&\lambda\\
      \frac 1{8\pi^2}&0
    \end{bmatrix}.
\end{equation}
The physical meaning of the real parameters $\lambda,g_{a,u}$ and $g_{a,d}$ will be explained later.

\begin{figure}[h]
  \begin{center}
\includegraphics[width=0.4\textwidth]{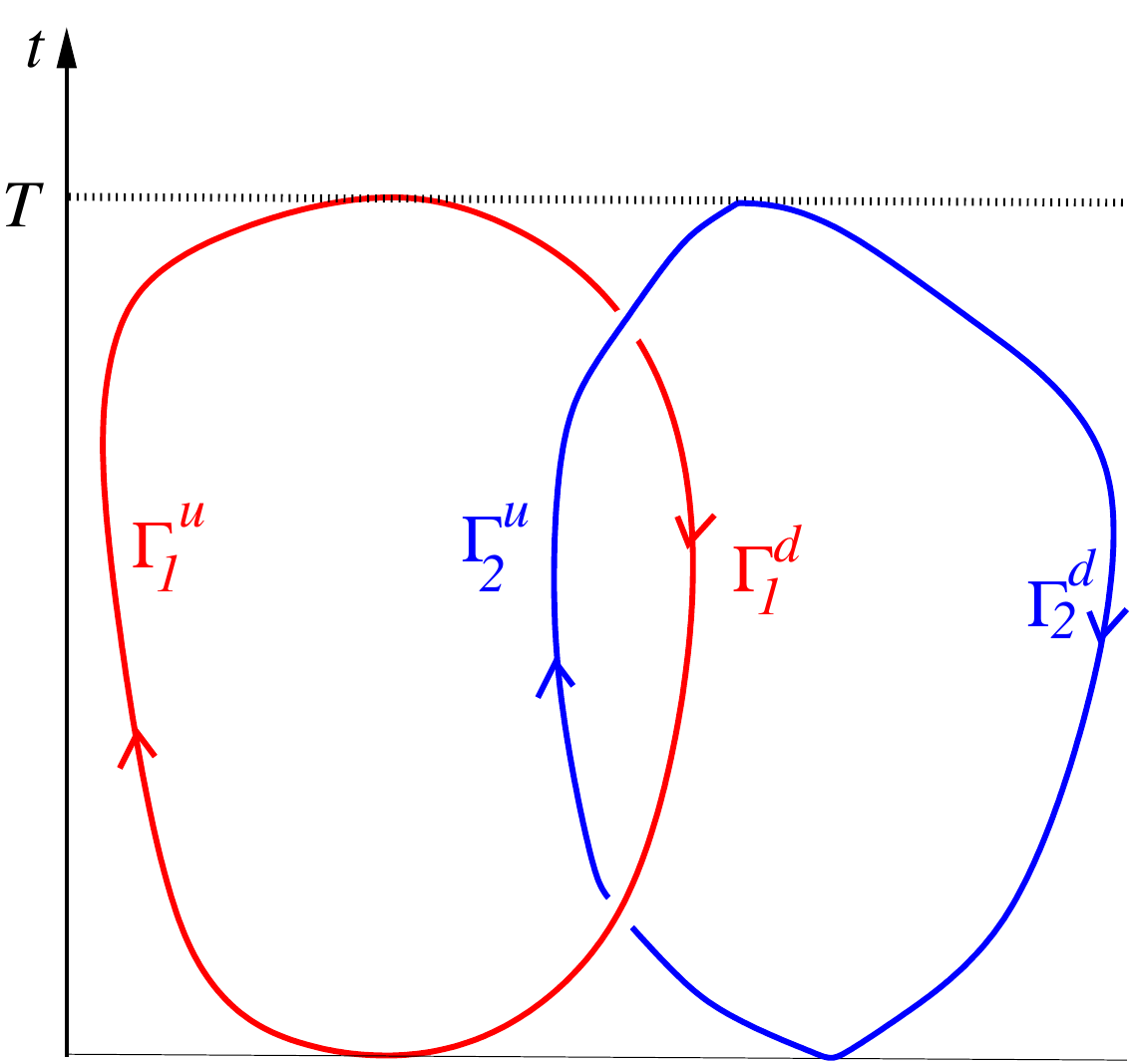}
  \end{center}
  \caption
      {
        Schematic picture of a $4-$plat realized by two concatenated polymer rings (ring 1 in read  and ring 2 in blue). In the $4-$plat configuration each ring $a$, with $a=1,2$, can be decomposed into two monotonic curves $\Gamma_a^u$ and $\Gamma_a^d$.
        The rings are oriented is such a way that  the heigth of $\Gamma_a^u$ always grows while that of $\Gamma_a^d$ decreases when going along the loop. The maximum height of both rings is $T$.
  }
  \label{figone}
  \end{figure}

It is straightforward to show that the following limit:
\begin{equation}
Z(\lambda)=\lim_{\alpha\to 0}\frac{Z(\lambda,\alpha)-Z(\lambda,0)}{\alpha}\label{pinttrue} 
\end{equation}
delivers the partition function $Z(\lambda)$ of Ref.~\cite{NAFFMP2024} that, in the limit of zero replicas, describes the statistical mechanics of two concatenated polymer rings, or [2]catenane,
whose topological constraints are imposed using the Gauss linking number.
An additional constraint is that the system should be in the conformation of a $4-$plat in which each ring has only a maximum and a minimum heights. An example of such a system is shown in Fig.~\ref{figone}.
Each ring $a=1,2$ consists of two monotonic curves $\Gamma_a^u$ and $\Gamma_a^d$.
The loops are oriented in such a way that the heights of $\Gamma_a^u$ are growing, while the heights of  $\Gamma_a^d$ are decreasing.
The rings are supposed to be in a solution at the so-called $\Theta-$point, so that the interactions between the monomers are absent.
As a consequence, the interaction terms $I_{sd}$ and $I_C$ are of entropic origin and arise due to the constraints.
This explains also why in Eq.~(\ref{pint}) the temperature is missing despite the fact that the $4-$plat fluctuates in a solution. The reason is that the model considered takes into account only the connectivity of the rings and the topological constraints due to the fact that they are concatenated. As a consequence, the internal energy of the system is always a constant and can always be set to zero. The upshot is that the free energy of the $4-$plat is given by $F=-\theta S$, where $S$ is the entropy and $\theta$ is the temperature. From this it follows that the partition function $Z(\lambda)=e^{-\beta F(\lambda)}=e^{-S/k_B}$, $k_B$ being the Boltzmann constant, is independent of the temperature.
The parameter $\lambda$ can be considered as the chemical potential for the Gauss linking number.  The $g_{a,u}$ and $g_{a,d}$ are the Kuhn lengths that are related to the flexibility of the lines $\Gamma_a^u$ and $\Gamma_a^d$.

The terms $I_{sd}$ and $I_C$ of Eqs.~(\ref{Isd}) and (\ref{Ic}) are related to the topological constraints.
$I_C$ takes into account the short-range interactions while the self-dual contribution $I_{sd}$ is responsible for the long-range interaction that are needed to preserve the topological state of the system.
It was noted in \cite{NAFFMP2024} that in the special case of homopolymer rings:
\begin{equation}
g_{a,u}=g_{a,d}=g\label{homop}
\end{equation}
the Coulomb-like interactions disappear. Moreover, in the limit of large values of the maximum height $T$, the contribution $I_T$ becomes negligible and the self-dual term $I_{sd}$ dominates.

At this point, we work under the homopolymer conditions (\ref{homop}) and the assumption that $T>>1$.  Thus, we restrict ourselves to the self-dual part of $S_{matter}$, in particular considering the equations defining the self-dual conformations:
\begin{eqnarray}
  \left(
  D_{a,1}^u+iD_{a,2}^u
  \right)\psi_a^{u(1)}(\boldsymbol x,t)&+&\alpha\psi_a^u(\boldsymbol r_{a,1},0)\delta( \boldsymbol x-\boldsymbol r_{a,2})\delta(T-t)\psi_c^{u*}(\boldsymbol r_{c,2},T)
\psi_c^u(\boldsymbol r_{c,1},0)
  \nonumber\\
  &\times&\prod_{b=1}^2\psi_b^{d*}(\boldsymbol r_{a,1},0)\psi_b^d(\boldsymbol r_{a,2},T)=0,\qquad\qquad c\ne a=1,2,\label{eqnreplone}
\end{eqnarray}
and
\begin{eqnarray}
  \left(
  D_{a,1}^u+iD_{a,2}^u
  \right)\psi_a^{u(\rho)}(\boldsymbol x,t)&=&0,\qquad\qquad\rho=2,\ldots,n.
\end{eqnarray}
Analogous equations may be written for the fields $\psi_a^{d}(\boldsymbol x,t)$
and for the complex conjugated fields $\psi_a^{u*}(\boldsymbol x,t),\psi_a^{d*}(\boldsymbol x,t)$.

We choose replica asymmetric solutions for which:
\begin{equation}
\psi_a^{u(\rho)}=\psi_a^{u(\rho)*}=\psi_a^{d(\rho)}=\psi_a^{d(\rho)*}=0,\qquad\qquad\rho=2,\ldots,n.
\end{equation}
The first replica fields $\psi_a^{u(1)},\psi_a^{u(1)*},\psi_a^{d(1)},\psi_a^{d(1)*}$ cannot be put to zero due to the term proportional to $\alpha$ in Eq.~(\ref{eqnreplone}). Otherwise, the partition function $Z(\lambda)$ of Eq.~(\ref{pinttrue}) will vanish identically and no degrees of freedom will be propagated.
Thus, we retain only the first replica fields, dropping the replica index for simplicity.
Since the parameter $\alpha$ can be considered very small due to the limit $\alpha\to0$, it is convenient to adopt the ansatz:
\begin{equation}
  \psi_a^{u,d}=\psi_{a,0}^{u,d}+\alpha\psi_{a,1}^{u,d}.
\end{equation}
Separating in Eq.~(\ref{eqnreplone}) the leading order contribution for the linear terms in $\alpha$, we obtain:
\begin{eqnarray}
  \left(
  D_{a,1}^u+iD_{a,2}^u
  \right)\psi_{a,0}^{u(1)}(\boldsymbol x,t)&=&0,\label{cond1}\\
\left(
  D_{a,1}^u+iD_{a,2}^u
  \right)\psi_{a,1}^{u(1)}(\boldsymbol x,t)&+&
  \alpha\psi_{a,0}^u(\boldsymbol r_{a,1},0)\delta( \boldsymbol x-\boldsymbol r_{a,2})\delta(T-t)\psi_{c,0}^{u*}(\boldsymbol r_{c,2},T)
\psi_{c,0}^u(\boldsymbol r_{c,1},0)
  \nonumber\\
  \times\prod_{b=1}^2\psi_{b,0}^{d*}(\boldsymbol r_{a,1},0)\psi_{b,0}^d(\boldsymbol r_{a,2},T)&=&0,\qquad\qquad c\ne a=1,2.\label{cond2}
\end{eqnarray}
It would be tempting to expand the partition function $Z(\lambda)$ around a self-dual solution of Eqs.~(\ref{cond1}-\ref{cond2}).
  Let's suppose that
  \begin{equation}
\eta_a^{u,d}=\psi_{a,0}^{u,d}+\alpha\psi_{a,1}^{u.d}
  \end{equation}
  is such a solution.
  Due to the fact that, if $T$ is large and we are working in the homopolymer case in which the conditions (\ref{homop}) hold, the self-dual term dominates the total action $S_{matter}$.  It is thus easy to show that, after taking the limit $\alpha\to0$ in Eq.~(\ref{pinttrue}), at the classical level:
  \begin{equation}
    Z(\lambda)=\prod_{a=1}^2\psi_{a,0}^{u*}(\boldsymbol r_{a,2},T)
    \psi_{a,0}^{u}(\boldsymbol r_{a,1},0)\psi_{a,0}^{d*}(\boldsymbol r_{a,1},0)
    \psi_{a,0}^{d}(\boldsymbol r_{a,2},T)\label{zlpt}
  \end{equation}
  As a matter of fact, $S_{matter}$ vanishes identically up to the first order in $\alpha$ included due to the equations (\ref{cond1}) and their analogues for the complex field configurations.
  The difficulty of this appealing approach based on the background field method, in which the partition function is expanded around a background configuration, is that there is a plethora of such configurations. For instance, it has been shown in \cite{NAFFMP2024} that there are several classes of ``static'' configurations that minimize the action $S_{matter}$, where static means independent of the height $t$. It is reasonable to assume that these configurations are not stable because of the thermal fluctuations that can easily make the system to pass from one  configuration to the other.
Nonetheless, the static solutions found in Ref.~\cite{NAFFMP2024} have quite peculiar properties, like for instance self-similarity and periodicities. It is thus licit to expect that all these solutions have some influence on the statistical mechanics of the two concatenated polymer rings if they are constrained to stay in $4-$plat conformations. As a check, we have performed numerical simulations, whose results will be presented in the next Section.
\section{From field theory to simulations}\label{numsim}
\begin{figure}[h]
  \begin{center}
\includegraphics[width=0.7\textwidth]{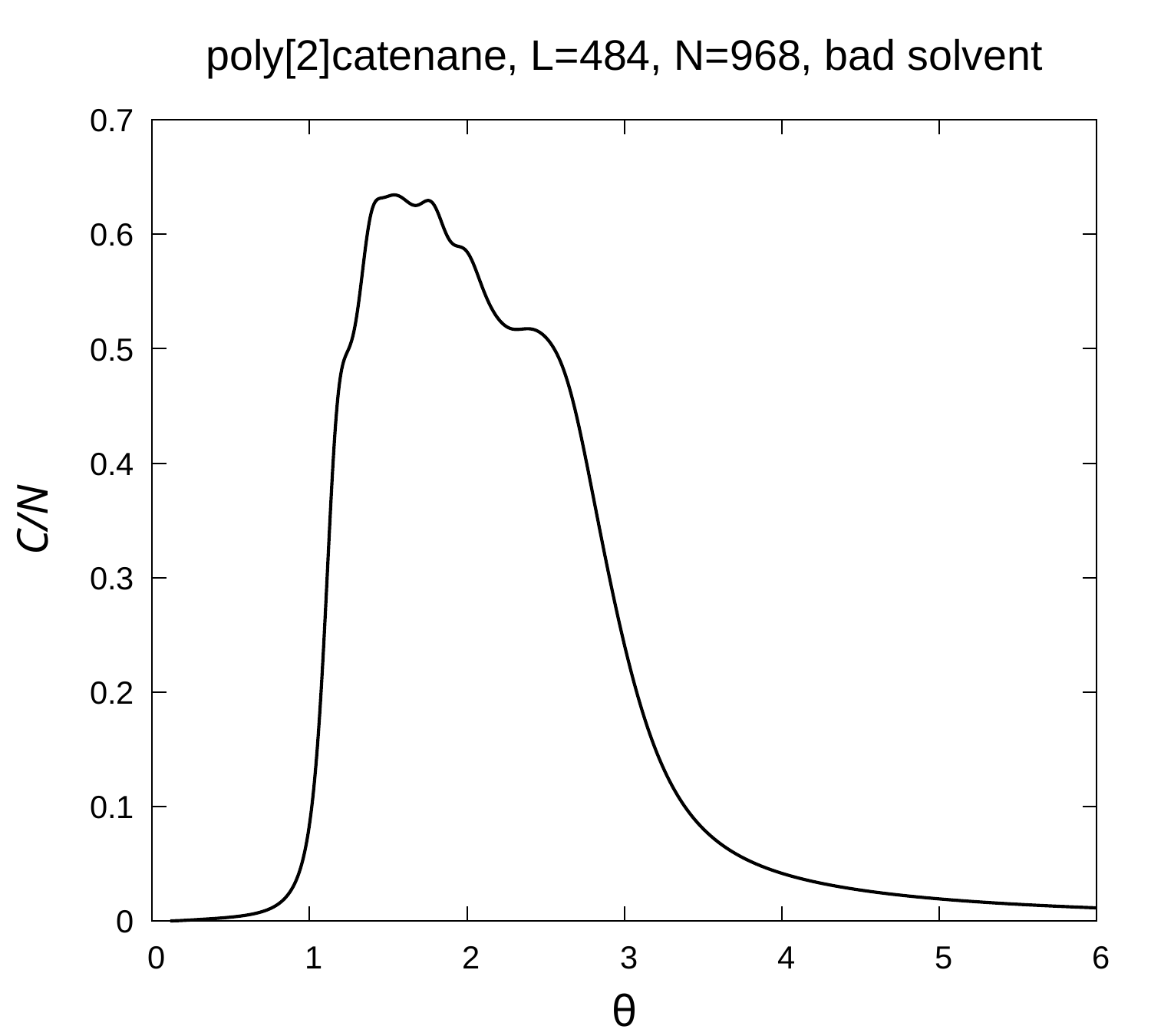}
  \end{center}
  \caption
{Plot of the specific heat capacity of a 2]catenane with $N=968$ monomers constrained to be in $4-$plat conformations.
  }
  \label{fig2}
  \end{figure}

\begin{figure}[h]
  \begin{center}
\includegraphics[width=0.7\textwidth]{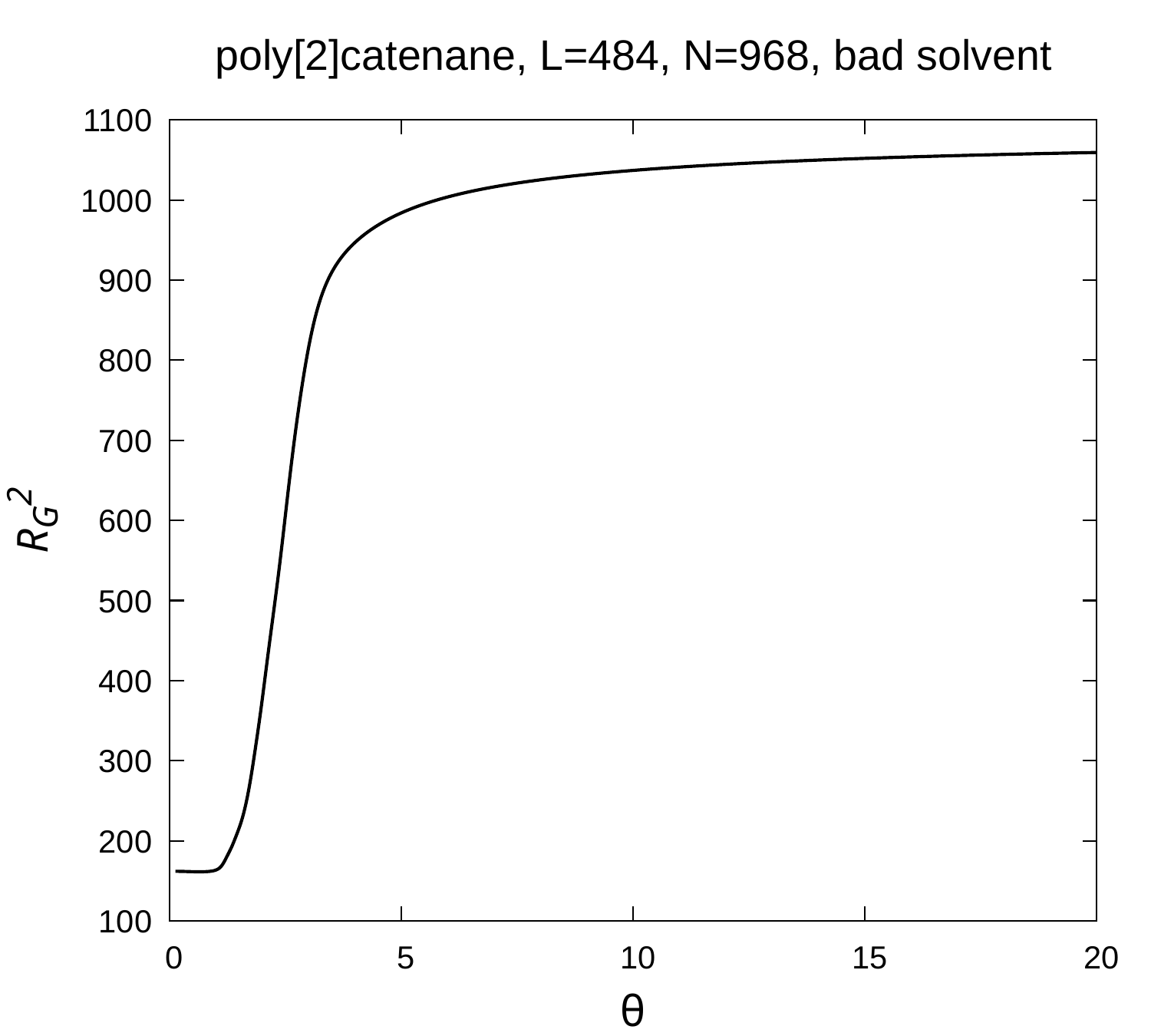}
  \end{center}
  \caption
{Plot of the mean square gyration radius $R_G^2$ of a [2]catenane with $N=968$ monomers constrained to be in $4-$plat conformations.
  }
  \label{2bis}
\end{figure}

\begin{figure}[h]
  \begin{center}
\includegraphics[width=0.7\textwidth]{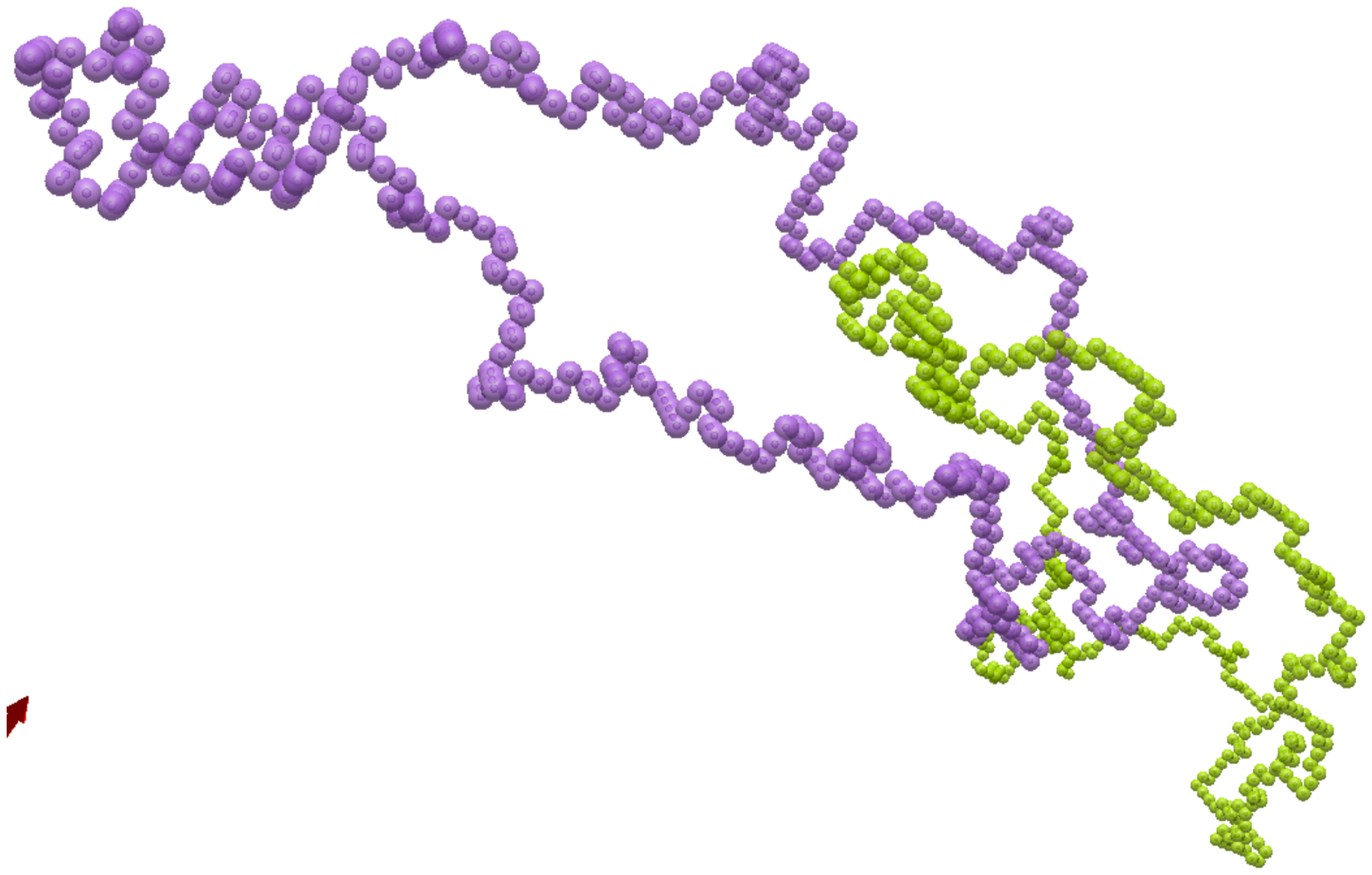}
  \end{center}
  \caption
      {        Plot of a conformation of the [2]catenane with specific energy $E=20$ (the energy is given here in dimensionless units). Such expanded conformations are common when temperatures are high and the thermal fluctuations are strong, so that the attractive interactions between the monomers become negligible. While the picture of the system has been rotated in order to fit it in a rectangle whose largest side is lying horizontally, it is worth noticing that the [2]catenane is elongated along the $t-$axis.}
  \label{fig3}
  \end{figure}

\begin{figure}[h]
  \begin{center}
\includegraphics[width=0.7\textwidth]{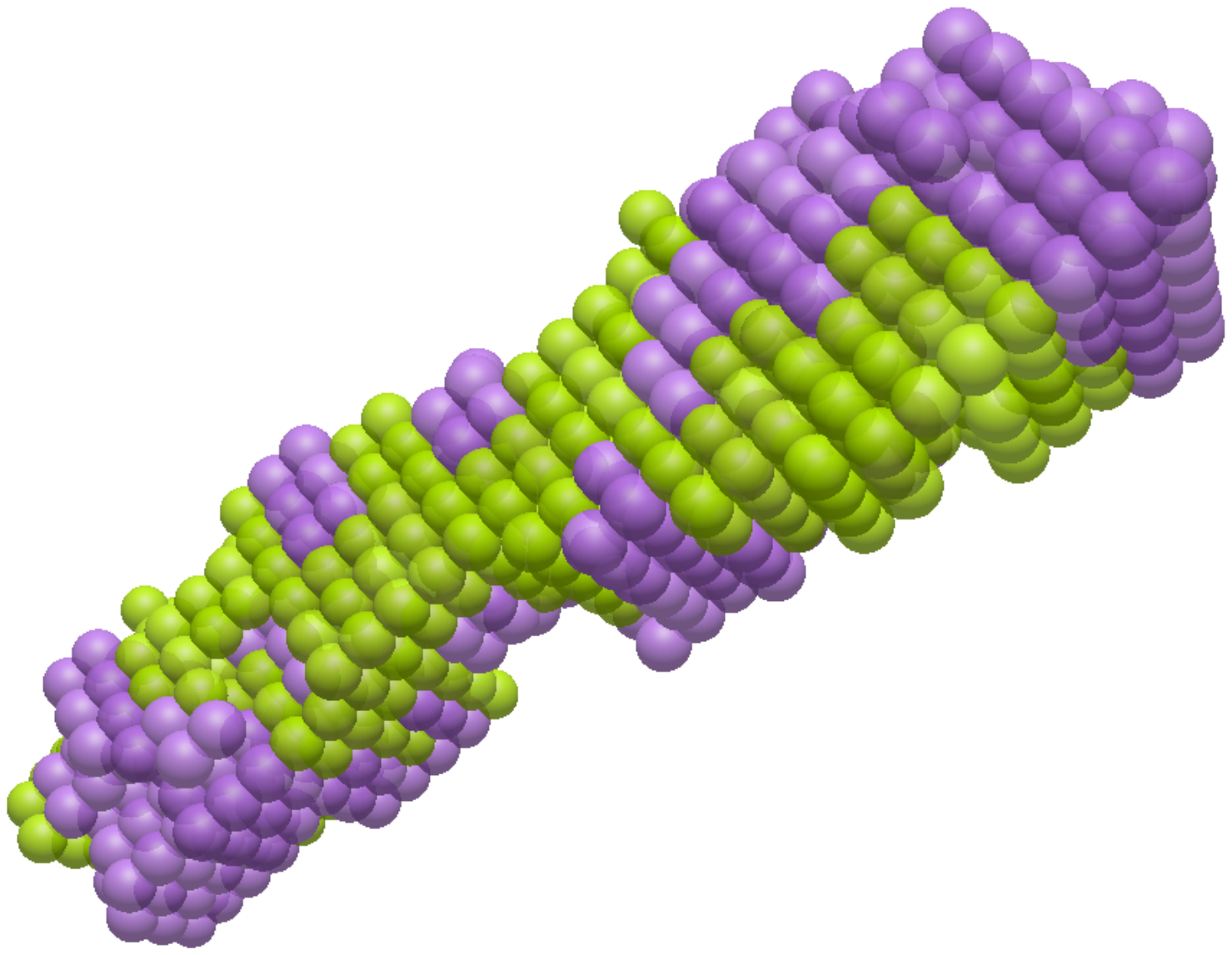}
  \end{center}
  \caption
      {Plot of a conformation of the [2]catenane with specific energy $E=1470$. Such conformations appear when the temperatures is very low (in this case $\theta=0.05$).
        Let us note that the [2]catenane is elongated along the $t-$axis.
  }
  \label{four}
  \end{figure}

\begin{figure}[h]
  \begin{center}
\includegraphics[width=0.7\textwidth]{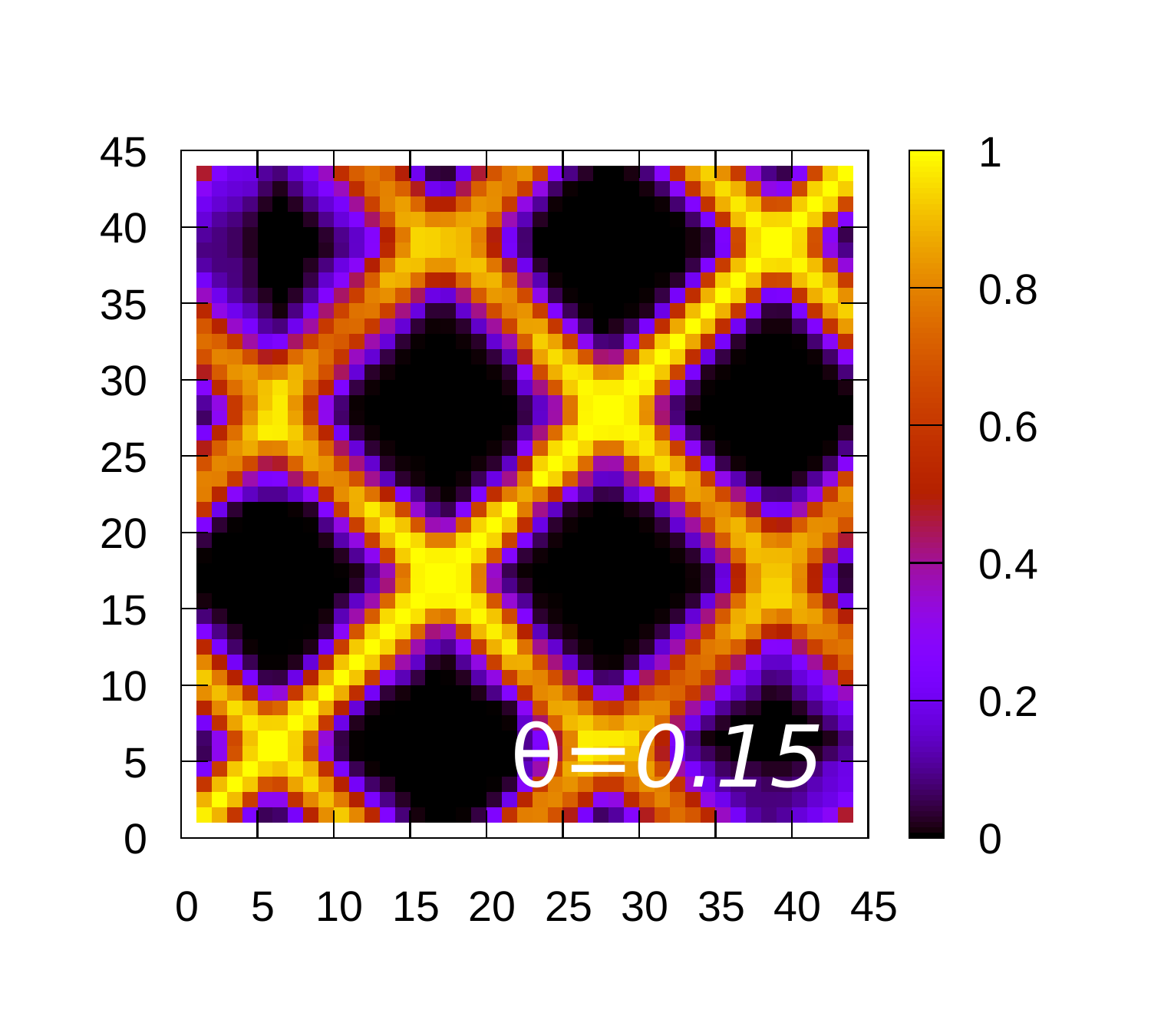}
  \end{center}
  \caption
      {Contact map of the $4-$plat at the low temperature $\theta=0.15$. Each ring has been divided into $22$ compartments. Each compartment contains 22 monomers.
        The contact map consists in a colormap containing a number of $44^2$ pixels.
        A pixel is identified by its position $\Sigma\Sigma'$ on the colormap, where $\Sigma$ numerates the rows and $\Sigma'$ the columns of the colormap. Each pixel $\Sigma\Sigma'$ expresses the probability that two compartments are in contact.
        Darker colors correspond to smaller probabilities.
        The probability is computed by averaging over several hundreds of billions of sampled conformations.
  }
  \label{six}
  \end{figure}

The simulations have been performed on a simple cubic lattice using a code described in Ref.~\cite{yzff} based on the Wang-Landau Monte Carlo algorithm~\cite{wl}.
The shortest distance on the lattice is $1$.
Starting from a seed conformation of the $4-$plat formed by two polymer rings with $L=484$ monomers each, the sampling is performed by applying  random transformations. The used random transformations are the pivot moves~\cite{madrasetal}.
The total number of monomers in the system is $N=968$.
The topological constraints are enforced by
  the pivot algorithm and excluded area (PAEA)  method of Ref.~\cite{yzff}. 
  Additionally, it is required that all conformations have only two points of minimum and two points of maximum.
  The minimum and maximum of ring~1 are located at the positions of the monomers 122 and 362 respectively, while the minimum and maximum of ring~2 are located at the positions of monomers 607 and 848 resepectively.
  The total heights of the rings are not preserved. 
  Another difference between the field theoretical model and the simulation is that in the latter short-range attractive interactions are allowed.
With a slight change of notation, now $\boldsymbol R_A$ is a three-dimensional
vector specifying the position of the $A$-th monomer, where $A=1,\ldots,968$.
The Hamiltonian of the system is given by:
\begin{equation}
H=-m_{AB}\epsilon\qquad\qquad A,B=1,\ldots,968,
\end{equation}
where
\begin{equation}
  m_{AB}=\left\{
    \begin{matrix}
      1&\mbox{if}&|\boldsymbol R_A-\boldsymbol R_B|=1\\
      0&&\mbox{otherwise}
    \end{matrix}
    \right.
\end{equation}
and $\epsilon$ is a positive energy scale.
The effects of the interactions become negligible at high temperatures due to the strong thermal fluctuations. As a consequence, in this regime the results of the simulations can be compared with those of the field theory, where interactions are absent.
The temperature will be denoted with the symbol $\theta$.
Finally, the rings are completely flexible and the segments connecting the monomers can be oriented in every direction without energy penalties. Of course, in the height direction the polymer lines must be monotonic.
The specific heat capacity $C/N$ of the system is shown in Fig~\ref{fig2}.
As it is possible to see, there is a single peak with bumps and a shoulder. The single peak is typical of knotted polymer rings in good and bad solvent, where there is a single phase transition from compact to extended coil phase, see e.g. \cite{yzff}.
With respect to single knots, however, the specific heat capacity exhibits an unusually broad peak,  because the expansion phase can be considered as concluded at $\theta\sim 4.00$.
For single knots, the transition stops already at $\theta\le 2$. The gyration radius of the $4-$plat jumps from $R_G^2\sim 170$ up to $R_G^2\sim 1067$ when the temperature grows from $\theta=0.05$ to $\theta=20$, see Fig.~\ref{2bis}.
In Fig.~\ref{fig3} it is displayed a conformation of the system that is typical of high temperatures, where the effects of the interactions become negligible due to the strong thermal fluctuations and the structure of the $4-$plat is purely entropic.
 Fig.~\ref{fig3} shows that the conformations of the two rings is elongated in one particular direction that actually coincides with the $t$ direction. 
This is a common characteristics of all the conformations of the system, independently of the temperature. For instance, an example of conformations appearing in the lowest studied temperature $\theta=0.05$ is provided in Fig.~\ref{four}.
Quite interestingly, at low temperatures  there is a periodicity in the way in which the two rings interact. This periodicity is not visible from Fig.~\ref{four} because it represents a single conformations, but becomes evident by looking at the contact maps that averages the contacts between the monomers over hundreds of billions of conformations.
To obtain the contact map of Fig.~\ref{six}, the system has been divided into 44 compartments containing 22 monomers each. A 45th compartment, which is coinciding with the first one, has been added for continuity,
The contact map consists of a colormap in which brighter colors correspond to a higher likelihood that  the compartments are in contact. By the definition used in this work, a contact is established when two compartment overlap, i.e., when the sum of their gyration radii is less than $1.5$ times the distance between their centers of mass.
Fig.~\ref{six} is naturally divided into four sectors.
The colormap is divided into $45^2$ pixels with pixel $\Sigma,\Sigma'$ corresponding to the interactions between the compartments $\Sigma$ and $\Sigma'$. The number of times in which all pairs of compartments enter in contact during a simulation is stored and out of this information the probability that they can be found in contact is computed.
Of course, going from left to right and from bottom to top, the upper and lower triangular components of the colormap are symmetric

The two sectors along the diagonal of the colormap in Fig.~\ref{six} can be considered as contact submaps that take into account the contacts of the compartments belonging to a single ring. The off-diagonal sectors correspond instead to the contact between compartments belonging to different rings.
As it is possible to see, the off-diagonal sectors are with good approximation repeating the dark-light patterns of the contact submaps in the diagonal sectors with a periodicity of 22 pixels. More precisely, let's assume that the compartments $\Sigma$ and $\Sigma'$  belong to ring $1$, Then,  $\Sigma +22$ and $\Sigma'+22$ are compartments of ring $2$.
From the colormap in Fig.~\ref{six}, it turns out that the probability that $\Sigma,\Sigma'$ are in contact is approximately equal to the probabilities that the compartments $\Sigma+22,\Sigma'$, $\Sigma,\Sigma'+22$ and $\Sigma+22,\Sigma'+22$ are in contact.
The only difference in the color patterns between the diagonal and off-diagonal sectors
(pairs $\Sigma+22,\Sigma'$ and $\Sigma,\Sigma'+22$, is that the pixels of the latter are slightly less bright. Moreover, there is a glitch in the color map forming a cross that divides the four sectors. This glitch is probably due to the fact that the two rings cannot be perfectly aligned along the $t$ axis due to the topological constraints.

\section{Conclusions}\label{conclusions}
In this work an alternative derivation of the field equation of motion of a $4-$plat studied in Ref.~\cite{NAFFMP2024} has been provided. This new derivation shows why, despite the zero replica limit, the degrees  of freedom corresponding to the first replica fields still propagate. In this way it has also been possible to compute  the explicit expression of the partition function of the $4-$plat on the background of any self-dual solution of the field equations of motion, see Eq.~(\ref{zlpt}).
Next, the system has been investigated using numerical simulations based on the Wang-Landau Monte Carlo algorithm.
In turns out that, at equilibrium, $4-$plats exhibit peculiar properties. First of all, despite the fact that both the points of minimum and maximum of the two rings are allowed to fluctuate, we have seen that the [2]catenane is organized in such a way that its structure is elongated along the special $t$ direction. Examples of these structures at different temperatures are provided in Figs.~\ref{fig3} and \ref{four}. Moreover, both rings have more or less the same height $T>>1$ and are aligned. This is exactly the situation considered in the field theoretical model. With respect to single knots, the transition from a compact state to the expanded coil state typical of polymers in a bad solvent takes place at a much higher temperature. Also the peak of the specific heat capacity of Fig.~\ref{fig2} is much broader. During the expansion, the gyration radius of the $4-$plat increases by a factor six as shown in Fig.\ref{2bis}.
There is also a symmetry at low temperatures between the two rings. Namely, the contact map of Fig.~\ref{six} is symmetric if a monomer $\Sigma$ of the first ring is replaced by the monomers $\Sigma+22$ of the second ring.

\begin{figure}[h]
  \begin{center}
\includegraphics[width=0.7\textwidth]{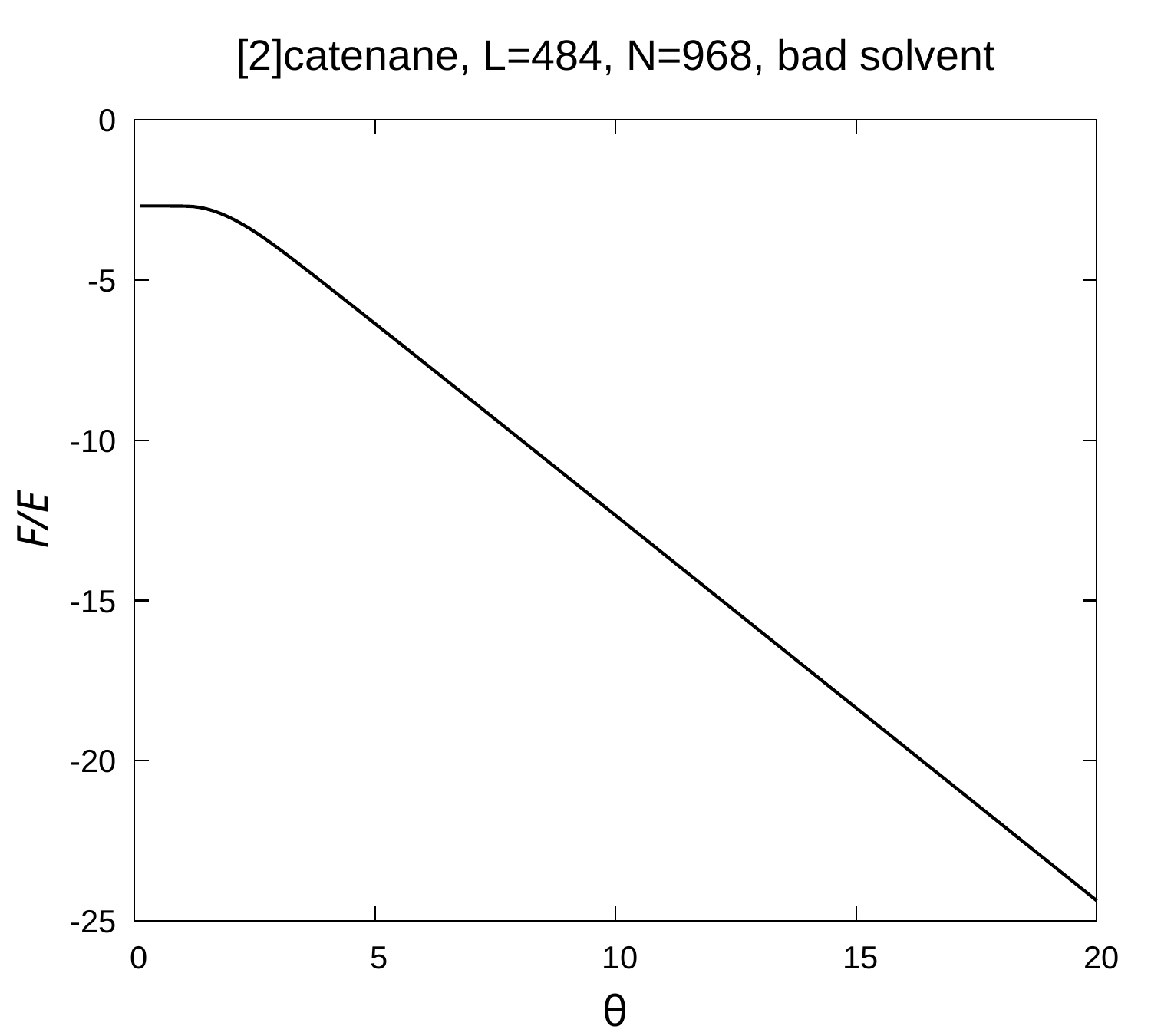}
  \end{center}
  \caption
      {Plot of the Helmholtz free energy $F/E$ of the [2]catenane with respect to the dimensionless temperature $\theta$.
  }
  \label{seven}
  \end{figure}

Finally, the field theoretical approach to $4-$plats has been compared with the results of numerical simulations obtained at high temperatures. The peculiar properties of the self-dual solutions  minimizing the static energy of the system derived in Ref.~\cite{NAFFMP2024} are not observed, apart from the already mentioned fact that the rings are aligned and their height is very large.
The absence of particular symmetries like those characterizing the analytical solution is probably due to the strong thermal fluctuations that make the system to pass from one of the many points of minimum in the complex energy landscape to the other.
Also the plot of the specific free energy of the system $F/E$ does not show points of minima, see Fig.~\ref{seven}.
  
\section*{Acknowledgments} 
The simulations reported in this work were performed in part using the HPC
cluster HAL9000 of
the University of Szczecin.
The research presented here has been supported by the Polish National Science Centre under
grant no. 2020/37/B/ST3/01471.
This work results within the collaboration of the COST
Action CA17139 (EUTOPIA).
The use of some of the facilities of the Laboratory of
Polymer Physics of the University of Szczecin, financed by 
a grant of the European Regional Development Fund in the frame of the
project eLBRUS (contract no. WND-RPZP.01.02.02-32-002/10), is
gratefully acknowledged.
Fruitful discussions with Luca Tubiana and Neda Abbasi Taklimi are gratefully acknowledged.



\begin{thebibliography}{1}%
\makeatletter
\providecommand \@ifxundefined [1]{%
 \@ifx{#1\undefined}
}%
\providecommand \@ifnum [1]{%
 \ifnum #1\expandafter \@firstoftwo
 \else \expandafter \@secondoftwo
 \fi
}%
\providecommand \@ifx [1]{%
 \ifx #1\expandafter \@firstoftwo
 \else \expandafter \@secondoftwo
 \fi
}%
\providecommand \natexlab [1]{#1}%
\providecommand \enquote  [1]{``#1''}%
\providecommand \bibnamefont  [1]{#1}%
\providecommand \bibfnamefont [1]{#1}%
\providecommand \citenamefont [1]{#1}%
\providecommand \href@noop [0]{\@secondoftwo}%
\providecommand \href [0]{\begingroup \@sanitize@url \@href}%
\providecommand \@href[1]{\@@startlink{#1}\@@href}%
\providecommand \@@href[1]{\endgroup#1\@@endlink}%
\providecommand \@sanitize@url [0]{\catcode `\\12\catcode `\$12\catcode
  `\&12\catcode `\#12\catcode `\^12\catcode `\_12\catcode `\%12\relax}%
\providecommand \@@startlink[1]{}%
\providecommand \@@endlink[0]{}%
\providecommand \url  [0]{\begingroup\@sanitize@url \@url }%
\providecommand \@url [1]{\endgroup\@href {#1}{\urlprefix }}%
\providecommand \urlprefix  [0]{URL }%
\providecommand \Eprint [0]{\href }%
\providecommand \doibase [0]{http://dx.doi.org/}%
\providecommand \selectlanguage [0]{\@gobble}%
\providecommand \bibinfo  [0]{\@secondoftwo}%
\providecommand \bibfield  [0]{\@secondoftwo}%
\providecommand \translation [1]{[#1]}%
\providecommand \BibitemOpen [0]{}%
\providecommand \bibitemStop [0]{}%
\providecommand \bibitemNoStop [0]{.\EOS\space}%
\providecommand \EOS [0]{\spacefactor3000\relax}%
\providecommand \BibitemShut  [1]{\csname bibitem#1\endcsname}%
\let\auto@bib@innerbib\@empty
\bibitem [{Note1()}]{Note1}%
  \BibitemOpen
  \bibinfo {note} {In the average, in a scale of time that is large with
  respect to the time scale of thermal fluctuations, this assumption is
  true.}\BibitemShut {Stop}%
\end{thebibliography}%


\begin{thebibliography}{999}
\bibitem{FF2004} F. Ferrari, {\it Topologically linked polymers are anyon systems}, {\it Phys. Lett. A} {\bf 323}(5-6) (2004), 351-359.
\bibitem{FFJPMPYZ2019} F. Ferrari, J. Paturej, M. Piatek, and Y. Zhao, {\it 	Knots, links, anyons and statistical mechanics of entangled polymer rings},
{\it Nucl. Phys. B} {\bf 945} (2019), 114673.
  \bibitem{ernst1} C. Ernst and D.W. Sumners, {\it A Calculus for Rational Tangles: applications
    to DNA recombinations}, {\it Math. Proc. Camb. Phil. Soc.} {\bf 108} (1990), 489-515.
\bibitem{ernst2}    C. Ernst and D.W. Sumners, {\it Solving Tangle Equations Arising in a DNA
  Recombination Model}, {\it Math. Proc. Camb. Phil. Soc.} {\bf 126} (1999), 23-36.
\bibitem{kaufmann1} L. H. Kauffman, S. Lambropoulou, {\it From Tangle Fractions to DNA}. In: Monastyrsky, M.I. (eds) {\it Topology in Molecular Biology. Biological and Medical Physics, Biomedical Engineering}, Springer, Berlin, Heidelberg, 2007.
\bibitem{sumners1} D. W. Sumners, {\it Knot Theory and DNA}, Proceedings of Symposia in Applied Mathematics {\bf 45} (1992), 39-72.
\bibitem{Vazquez}  M. Vazquez and D. W. Sumners, {\it Tangle Analysis of Gin Site Specific Recombination}, {\it Math. Proc. Camb. Phil. Soc.} {\bf 136} (2004), 565-582
  \bibitem{Vazquez2} N. R. Beatona, K. Ishiharab, M. Atapour,  J. W. Eng,, M. Vazquez,
K. Shimokawah, and Ch. E. Soteros, {\it Entanglement statistics of polymers in a lattice tube and
  unknotting of $4-$plats'}, arXiv:2204.06186v3.
\bibitem{fujita}
  M. Fujita, {\it Self-assembly of [2] catenanes containing metals in their backbones}, {\it Accounts of chemical research}, {\bf 32}(1) (1999), 53-61.
\bibitem{wassermann1}  E. Wasserman, {\it J. Am. Chem. Soc.} {\bf  82} (1960), 4433 – 4434;
\bibitem{wassermann2} H. L. Frisch, E. Wasserman, {\it Chemical Topology}. {\it J. Am. Chem. Soc.}{\bf 83} (19610, 3789-3795.
\bibitem{Alex1} D. Becerra, A. R. Klotz, and L. M. Hall, {\it Single-molecule analysis of solvent-responsive mechanically interlocked ring polymers and the effects of nanoconfinement from coarse-grained simulations}, {\it Journal of Chemical Physics}, {\bf 160}(11) (2024), 114906.
\bibitem{Alex2}   A. R. Klotz, {\it 
  Borromean hypergraph formation in dense random rectangles},
  {\it Physical Review E} {\bf 110}(3) (2024), 034501.
  \bibitem{Mesfin} H. Guo, K. Qian, and M. Tsige,  {\it Theta Temperature Depression of Mechanically Interlocked Polymers: [2]catenane as a Model Polymer}, {\it Macromolecules, } {\bf 56}(22) (2023), 9164-9174.
  \bibitem{Eutopiareview} L. Tubiana, G. P Alexander, A. Barbensi, D. Buck, J. HE Cartwright, M. Chwastyk, M. Cieplak, I. Coluzza, S. Copar, D. J Craik, M, Di Stefano, R. Everaers, P. FN Faísca, F. Ferrari, A. Giacometti, D. Goundaroulis, El. Haglund, Y.-M. Hou, N. Ilieva, S. E Jackson, A. Japaridze, N. Kaplan, A. R Klotz, H. Li, C. N Likos, E. Locatelli, T. López-León, T. Machon, Cr. Micheletti, D. Michieletto, A. Niemi, W. Niemyska, S. Niewieczerzal, F. Nitti, E. Orlandini, S. Pasquali, A. P. Perlinska, R. Podgornik, R. Potestio, N. M. Pugno, M. Ravnik, R. Ricca, C. M Rohwer, A. Rosa, J. Smrek, A. Souslov, A. Stasiak, D. Steer, J. Sułkowska, P. Sułkowski, De Witt L Sumners, C. Svaneborg, P. Szymczak, T. Tarenzi, R. Travasso, P. Virnau, D. Vlassopoulos, P. Ziherl, S. Zumer, {\it Topology in soft and biological matter}, {\it Phys. Rep. } {\bf 1075} (2024), 1-137.
  \bibitem{Rensburg} E. J. Janse van Rensburg, E. Orlandini, M. C. Tesi and S. G. Whittington, {\it J. Phys. A: Math. Theor.} {\bf 55} (2022), 435002.
  \bibitem{NAFFMP2024} N. Abbasi Taklimi, F. Ferrari, and M. R. Pia\c{a}tek,
    {\it Self-dual solutions of a field theory model of two linked rings},
    {\it Nuclear Physics B} {\bf 999} (2024), 116447.
    \bibitem{Dunne} G. Dunne, {\it Self-Dual Chern-Simons Theories}, Springer Science \& Business Media, 2009.
    \bibitem{abbott} L. F. Abbott, {\it Introduction to the background field method}. {Acta Phys. Pol. B} {\bf 13} (1981) (CERN-TH-3113), 33-50.
\bibitem{wl}F. Wang and D. P. Landau, Phys. Rev. Lett. 86 (2001), 2050.
\bibitem{yzff} Y. Zhao and F. Ferrari, JSTAT {\it J. Stat. Mech.}
(2012), P11022.
\bibitem{madrasetal}  N. Madras, A. Orlistsky and L. A. Shepp, Journal of
  Statistical Physics 58, 159 (1990).
\end{thebibliography}
\end{document}